\documentclass{caosp}

\usepackage{graphicx}

\articleNo{123}
\pubyear{2005}
\volume{35}
\volnumber{3}
\firstpage{1}
\received{May 1, 2005}
\accepted{August 28, 2005}

\newcommand{\aand}{$\alpha$~And}
\newcommand{\cam}{53~Cam}
\newcommand{\cvn}{$\alpha^2$~CVn}

\begin{document}

\htitle{Magnetic fields, spots and weather in chemically peculiar stars}
\hauthor{O. Kochukhov}

\title{Magnetic fields, spots and weather in \\ chemically peculiar stars}


\author{O. Kochukhov}

\institute{
Department of Astronomy and Space Physics, 
Uppsala University, \\ Box 515, SE-751 20 Uppsala, Sweden
}

\date{December 1, 2007}

\maketitle

\begin{abstract}
New observational techniques and sophisticated modelling methods has led to dramatic
breakthroughs in our understanding of the interplay between the surface
magnetism, atomic diffusion and atmospheric dynamics in chemically peculiar
stars. Magnetic Doppler images, constructed using spectropolarimetric
observations of Ap stars in all four Stokes parameters, 
reveal the presence of small-scale field topologies.
Abundance Doppler mapping has been perfected to the level where 
distributions of many different chemical elements can be deduced self-consistently
for one star.
The inferred chemical spot structures are diverse and do
not always trace underlying magnetic field geometry. Moreover, horizontal chemical
inhomogeneities are discovered in non-magnetic CP stars and evolving chemical
spots are observed for the first time in the bright mercury-manganese
star \aand. These results show that in addition to magnetic fields,
another important non-magnetic structure formation mechanism acts in CP
stars.
\end{abstract}

\section{Introduction}

Chemically peculiar stars are privileged astrophysical laboratories for
investigation into the structure formation processes taking place in the
surface layers of stars. Unlike the turbulent atmospheres of cool solar-type
stars or mass loss-dominated envelopes of massive stars, atmospheres of slowly
rotating main sequence A and B stars are very stable, and this in turn allows
several remarkable processes to operate in these stars. The most prominent
characteristic of many CP stars is the presence of strong fossil magnetic
fields at their surfaces and in interiors. Moreover, due to the atmospheric
stability, further enhanced by the magnetic field, chemical segregation processes,
determined by the competition between gravitational settling and radiative
levitation, can operate efficiently. These processes are responsible for conspicuous
non-solar chemical abundance  patterns observed in CP stars and can also lead to
strong vertical abundance gradients in the line-forming regions. The balance
between gravitation and radiation pressure is affected by the magnetic field and
hence prominent horizontal chemical inhomogeneities emerge at the surfaces of
CP stars. Rapid pulsational variability in cool Ap stars, also related to
non-solar chemistry and the presence of strong magnetic field, adds dynamical
aspect to the intricate picture of the CP-star phenomenon. 

Various manifestations of this unique and poorly understood combination of
complex surface properties of CP stars are readily observed in the form of
periodic light and spectrum variation, anomalous line profile shapes,
significant Zeeman broadening, circular and linear polarization in spectral
lines. With some effort these observables can be interpreted in terms of local
magnetic field strength and orientation, horizontal and vertical abundance
maps, and time-dependent pulsation velocity field. The resulting extensive information on
the surface structures and atmospheric dynamics  is often utterly unique and is usually
unaccessible for other kinds of stars, making CP stars the only source of
crucial direct evidence and powerful observational constraints on the physical
processes governing chemical diffusion, emergence and evolution of large-scale
magnetic field. 

At the same time, the complexity and remarkable diversity of the phenomena
observed in the atmospheres of CP stars drive the development of new
astronomical instrumentation, modelling techniques and computer codes. On this
background continuing efforts of many CP-star specialists, with enthusiastic
support from the rest of the stellar community, has led to the revival of this
research field in the last decade. New discoveries have overturned some of the old
paradigms, generally resulting in a more complex, yet also more interesting
picture of chemically peculiar stars. 

In this review I highlight recent progress achieved in the studies of magnetic fields
and horizontal chemical inhomogeneities in CP stars. The closely related topics
of chemical stratification and magnetoacoustic pulsations in roAp stars are
dealt with in other contributions (Ryabchikova, Sachkov et al., this meeting).

\section{Magnetic fields}

Multipolar fits to the phase curves of magnetic observables (Landstreet \&
Mathys \cite{LM00}; Bagnulo et al. \cite{BLL02}) is a common technique to infer
information about the large-scale structure of the magnetic fields in CP stars.
Reasonably successful application of this modelling method to many stars over the
last few decades is often taken as an indication that magnetic topology in most CP
stars is close to a centered dipolar structure, sometimes with an offset
or a small quadrupolar contribution. However, one cannot guarantee that
this conclusion is not an artifact of the {\it a priori} low-order multipolar field
assumption, applied to often quite sparse set of field modulus and longitudinal
magnetic measurements. Indeed, a number of studies that attempted to incorporate
information on linear polarization in multipolar modelling (Leroy et al.
\cite{LLL96}; Bagnulo et al. \cite{BWD01}) have demonstrated that significant
deviations from the dipolar field geometry are present in many stars. Another concern is
fundamental non-uniqueness of multipolar modelling, which sometimes yields widely
different results, depending on what particular type of several common field
parameterizations is adopted (Kochukhov \cite{K06}). Thus, multipolar fit to the
phase curves of magnetic observables is generally unable to give direct information
on the stellar magnetic field topology. It only characterizes the basic
statistical properties (the mean and few lowest moments) of the surface distribution
of magnetic field modulus and the line of sight field component.

The ultimate breakthrough in answering the question how does the CP-star magnetic
field really look like was achieved with the help of high-resolution four Stokes
parameter observations, secured by Wade et al. (\cite{WDL00}). They have obtained
measurements of circular and linear polarization in the profiles of individual metal
lines for several bright Ap stars using the MuSiCoS spectropolarimeter 
at the Pic-du-Midi Observatory. These data are being interpreted with the
Magnetic Doppler imaging (MDI) method. Developed by Piskunov \& Kochukhov
(\cite{PK02}) and thoroughly tested by Kochukhov \& Piskunov (\cite{KP02}), this new
implementation of Doppler imaging allows one to obtain self-consistent maps of
magnetic field and chemical spots through the regularized inversion of the 
rotationally modulated Stokes $IQUV$ time series. This technique makes possible a
robust reconstruction of the magnetic maps without any {\it a priori} (e.g. multipolar)
assumptions about magnetic field geometry. 

\begin{figure}[!t]
\centering
\includegraphics[width=\hsize]{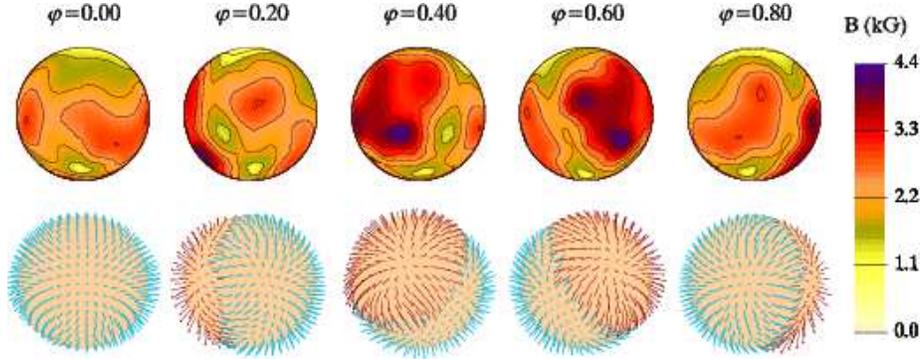}
\caption{Magnetic field topology of \cvn\ inferred with the help of magnetic Doppler
imaging in all four Stokes parameters (Kochukhov \& Wade,
in preparation). The upper row of spherical plots shows distribution of the 
field strength, the lower row presents the map of field orientation. 
The star is shown 
at the inclination angle $i=120^{\rm o}$ for five rotation phases.}
\label{fig1}
\end{figure}

Kochukhov et al. (\cite{KBW04}) applied MDI to the MuSiCoS observations of \cam,
constructing the first stellar magnetic map based on observations in all four Stokes
parameters. The outcome of the magnetic inversion for \cam\ was rather unexpected: it was
found that this Ap star hosts surprisingly complex magnetic field, which cannot be
approximated by any low-order multipolar topology. Subsequently we have carried out
MDI with the MuSiCoS four Stokes parameter spectra of another well-known CP star,
\cvn\ (Kochukhov \& Wade, in preparation). The result of the magnetic inversion from the
Cr and Fe lines is presented in Fig.~\ref{fig1}. Unlike \cam, \cvn\ shows a more
subtle deviation from the dominant dipolar field configuration. Nevertheless, the
small-scale features, such as the two magnetic spots visible at phase
$\varphi\approx0.5$, are required to reproduce variation of  the Stokes parameter
spectra. Remarkably, these fine details of the surface magnetic field distribution
could be inferred only when Stokes $Q$ and $U$ profiles are employed for magnetic
mapping. The highest quality observations of traditional integral magnetic
observables, aided with the high-resolution Stokes $I$ and $V$ spectropolarimetry, 
apparently do not contain information sufficient to resolve the small-scale magnetic
topologies (Kochukhov et al. \cite{KPI02}; Khalack \& Wade \cite{KW06}; L\"uftinger
et al., this meeting). It is therefore conceivable that the typical field structure of
magnetic CP stars contains contributions both from a global, dipolar-like component
and from much smaller spatial scales. 

The small number of stars analysed up to now with the MDI technique leaves open the
question of short-term stability of the newly discovered complex magnetic elements,
as well as possible evolutionary and star-to-star scatter in the relative
importance of large vs. small-scale magnetic structures. Evidently, repeated mapping of
selected CP stars and substantial expansion of the sample of stars with detailed
magnetic maps is needed. This is becoming feasible thanks to the recent 
commissioning of the two new spectropolarimeters, NARVAL and ESPaDOnS, installed at
Pic-du-Midi and CFHT. Compared to MuSiCoS, these instruments tremendously improve the
$S/N$, wavelength coverage and resolution of the Stokes parameter spectra. We are
using both spectrographs in a long-term observing campaign (Silvester et al., this
meeting) to collect observations needed for exploring in detail the field topology at
various spatial scales and for probing intrinsic changes of the small-scale field
structures.

\section{Chemical spots}

Since the formulation of oblique rotator model (Stibbs \cite{S50}) and the
first attempts to determine distribution of chemical elements based on the
rotational modulation of radial velocity and equivalent widths of variable
spectral lines (Deutsch \cite{D58}), the studies of chemical non-uniformities
have become an important part of CP-star research. Inspired by the phenomenon of
chemical spots, the powerful technique of Doppler imaging (DI) was developed
for CP stars (Goncharskij et al. \cite{GSK82}) and then extended to the problems
of mapping temperature inhomogeneities, magnetic fields and pulsation velocity
field. 

\begin{figure}[!t]
\centering
\includegraphics[width=11cm]{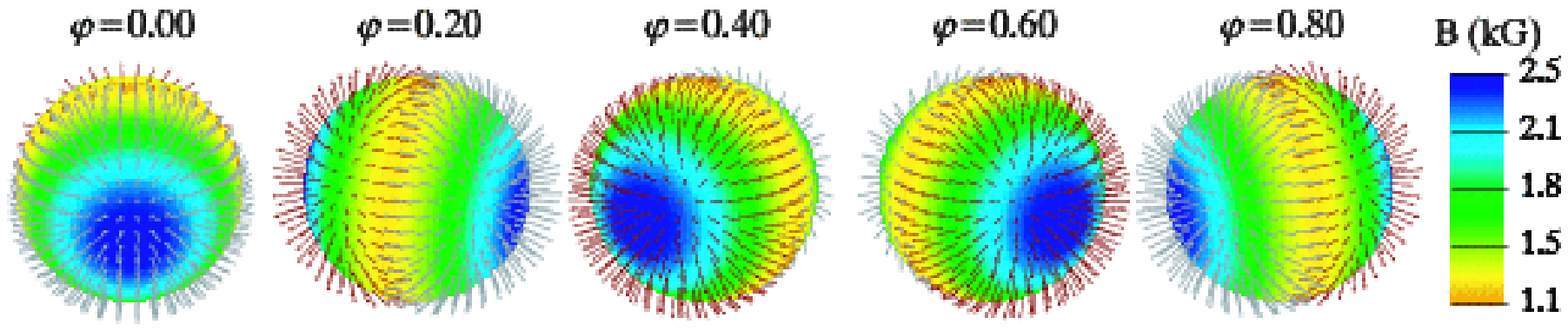}
\includegraphics[width=11cm]{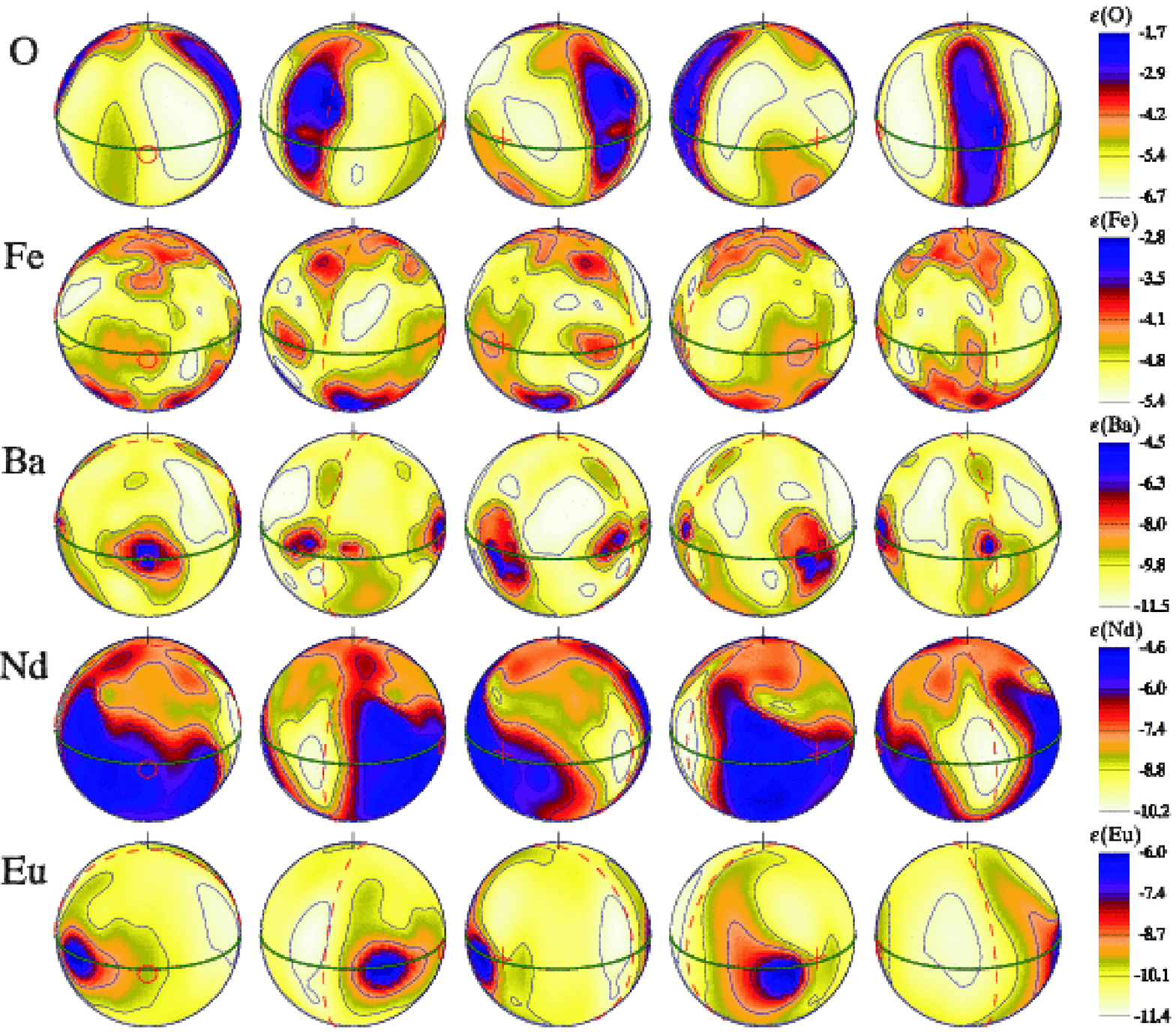}
\caption{Surface magnetic and chemical structure of  
the roAp star HD\,83368. The upper row shows dipolar magnetic field topology
of the star. Five lower rows present surface distribution of O, Fe, Ba, Nd
and Eu (Kochukhov et al. \cite{KDP04}). In each spherical plot the star is
shown at the inclination angle $i=68^{\rm o}$ for five rotation phases. 
}
\label{fig2}
\end{figure}

Over the years the DI technique was repeatedly applied to study spots of individual
elements in selected stars (e.g., Rice et al. \cite{RWH97}; Hatzes \cite{H97}). However,
the emerging recent trend is to perform multi-element abundance DI studies, covering large
number of elements in a self-consistent manner. Such spatially-resolved abundance analyses
provide a more representative and comprehensive picture of the surface chemical
inhomogeneities. Ongoing investigations, pursuing multi-element mapping of cool Ap stars
HD\,24712 (L\"uftinger et al.) and HD\,3980 (Obrugger et al.), were presented at this
meeting. Another examples of such studies can be found in L\"uftinger et al.
(\cite{LKP03}) for $\varepsilon$~UMa and in Kochukhov et al. (\cite{KDP04}) for the
prototype rapidly oscillating Ap star HD\,83368. Five of the 17 abundance maps
reconstructed in the latter study are presented in Fig.~\ref{fig2}. Chemical maps shown in
this plot illustrate an extremely wide range of behaviour observed for different elements
in the same star. Some species, for instance neutral oxygen, show global features,
obviously related to the symmetry of the dipolar magnetic field component. Other elements, for
example Fe and Ba, are distributed in a considerably more complex surface pattern, defying
a simple interpretation in terms of underlying magnetic field geometry.

Another important feature of modern multi-element DI maps is the discrepancy in
the surface distribution of elements with similar atomic weights, in particular
different rare-earth elements (REEs). Old DI studies of REE spots in hot
magnetic stars (e.g. Goncharskij et al. \cite{GRS83}), typically limited to Eu,
have inferred overabundance of this element at the magnetic poles. It is not
uncommon to meet an overly simplistic and misleading generalization of these
results to all REEs in all Ap stars. Modern DI studies have probed
distributions of REEs other than Eu (Kochukhov et al. \cite{KPI02,KDP04}) and
have clearly rejected the notion of a common surface structure of heavy
element spots. As seen on the example of HD\,83368 (Fig~\ref{fig2}), Eu spots
are small and high-contrast, but are displaced from the magnetic poles of that
particular star. At the same time, Nd and Pr, responsible for many strong
variable lines in the Ap-star spectra, exhibit markedly different surface
distribution, dominated by the ring of relative underabundance rather than a clearly
defined polar spot.

The diversity and occasional extreme complexity of the chemical maps derived in recent
studies of spotted Ap stars suggest that chemical diffusion operates in different regimes
for different elements and ions. For some species radiative forces, modulated by the
magnetic field, are overwhelming and create large-scale features. For other elements
radiative acceleration may be more finely balanced by gravity and hence the observed local
abundance may be easily perturbed by small local variations of magnetic field and atmospheric
structure. Moreover, the presence of chemical overabundances at the intersection of
magnetic and rotational equators (cf. Ba in HD\,83368) hints that magnetically-channeled
mass loss plays a non-negligible role in the process of chemical spot formation.

The question of possible time variation of the chemical spot geometry in magnetic stars has
not received sufficient attention. The light curve stability of the photometrically variable CP
stars (Adelman \& Kaewkornmaung \cite{AK05}) and a sequence of Si Doppler maps produced for
56~Ari (Ryabchikova \cite{R03}) show no evidence of surface structure changes over the
time span of  several decades. However, only the largest spatial scales 
are  probed in these studies and, therefore, no data can currently exclude evolution of
the small-scale abundance structures resolved in modern DI studies.

At the same time, it is becoming clear that under the right circumstances radiative
diffusion can produce chemical spots rather quickly compared to the stellar evolutionary
time scales. The search for progenitors of Ap/Bp stars among the Herbig Ae/Be objects
(Wade et al. \cite{WBD07}) has revealed a number of infant magnetic CP stars. For one such
object -- the B9 primary of the HD\,72106 system -- Doppler mapping carried out by Folsom
et al. (this meeting) shows well-developed typical Ap-star abundance spots, despite the
fact that the star has completed less than 1.5\% of its main sequence life, and possibly
still has not reached ZAMS.

\section{Chemical weather}

The classification scheme of the upper main sequence CP stars has traditionally
relied on the dichotomy between the magnetic and non-magnetic peculiar stars
(Preston \cite{P74}). The former (Ap/Bp stars) possess strong organized
magnetic fields, inhomogeneous horizontal abundance distributions, and show
large-amplitude photometric and spectroscopic variation. The latter (Am and
HgMn stars) were thought to be constant and lack magnetic fields and chemical
spots, sharing with their magnetic counterparts only the unusually slow
rotation. Repeated searches of magnetic fields in Am and HgMn stars have failed
to produce any detection of a large-scale Ap-like or small-scale solar-like
field, even when the highest quality spectropolarimetric data were used and the
errors of the effective field measurements were brought down to 30--50~G
(Shorlin et al. \cite{SWD02}). Rare reports of the photometric variability of Am
and HgMn stars were always refuted with higher quality data (Adelman
\cite{A93}), while claims of marginal line profile changes (Malanushenko
\cite{M96}; Zverko et al. \cite{ZZK97}) were not considered convincing.

The observational situation has changed radically with the discovery of mercury
abundance inhomogeneities in the brightest HgMn star \aand\ (Adelman et al.
\cite{AGK02}). Careful observations of this broad line star have demonstrated
large-amplitude variability of the Hg~{\sc ii} 3984~\AA\ line and
allowed to measure rotation period of a HgMn star for the first time. The mercury
line variability, established beyond any doubt, was interpreted with a Doppler
imaging code. The resulting map showed a series of high-contrast mercury spots,
located along the rotational equator. 

This remarkable discovery of the new type of spotted structure has questioned
our understanding of the basic properties of HgMn stars and their
differentiation from the magnetic CP stars. One can speculate that the stellar
properties change continuously from strongly magnetic Bp to HgMn stars and
that weak magnetic fields, present in the latter class, could be responsible
for the chemical inhomogeneities in \aand. However, this hypothesis did not
withstand the test of very high-precision spectropolarimetric observations of
\aand\ carried out by Wade et al. (\cite{WAB06}), who proved that no line of
sight field stronger than 10~G can be detected in this star. The dipolar field
strength in the atmosphere of \aand\ is thus constrained to be below the
equipartition limit of $\approx$\,250~G. Such field is unlikely to affect
chemical diffusion and, even if present at the epoch of star formation, is
expected to decay rapidly due to shearing by the differential rotation
(Auri\'er et al., this meeting).

It was soon demonstrated that the spots on \aand\ are not unique. Kochukhov et al.
(\cite{KPS05}) have surveyed a small group of HgMn stars with $T_{\rm eff}$ and
$v\sin i$ close to those of \aand, finding signatures of spotted Hg
distribution in two (HR\,1185, HR\,8723) out of five stars studied. Hubrig et
al. (\cite{HGS06}) have confirmed tentative finding of Zverko et al.
(\cite{ZZK97}) that the Hg~{\sc ii} 3984~\AA\ line in the primary component of
the eclipsing SB2 HgMn binary star AR~Aur shows intrinsic changes. Furthermore, Hubrig
et al. also reported variability of Y, Zr and Pt in AR~Aur~A. Our independent
spectropolarimetric observations of this system have confirm these results and
also showed that no global magnetic field is present on
either component of AR~Aur.

\begin{figure}[!t]
\centering
\includegraphics[width=11cm]{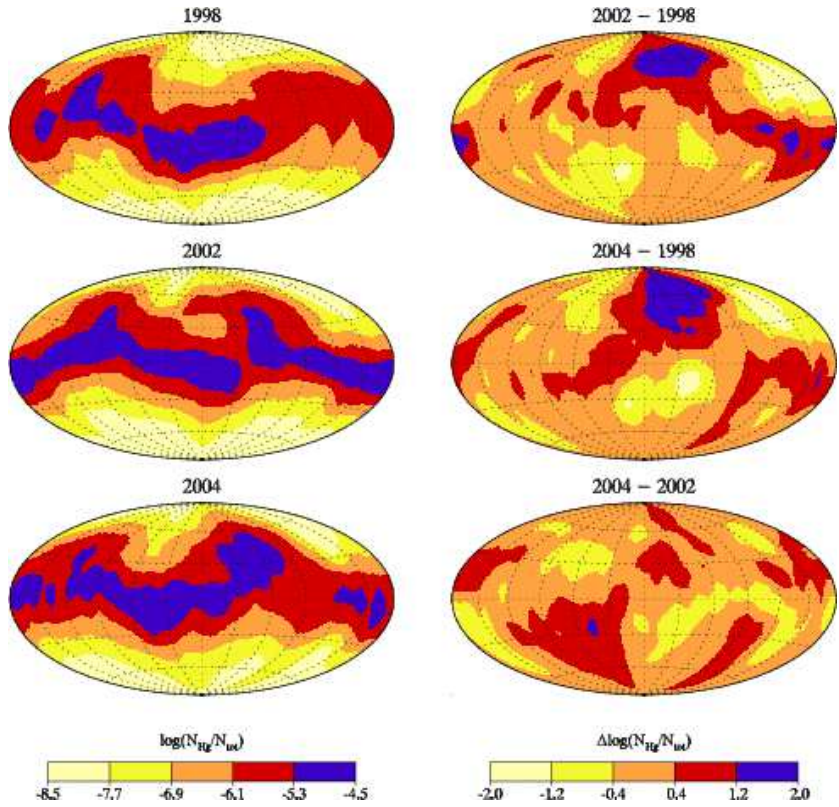}
\caption{Evolution of mercury inhomogeneities in the atmosphere of HgMn star \aand\
(Kochukhov et al. \cite{KAG07}).
This figure shows Hammer-Aitoff projection of Hg maps for three different epochs 
covering the time 
span of 7 years. {\it Left column:} abundance distributions for years 1998, 2002
and 2004. {\it Right column:} pairwise difference of the Hg maps. 
}
\label{fig3}
\end{figure}

Thus, it appears that chemical inhomogeneities are widespread among HgMn stars
and that previous failures to notice them are largely related to a small number of
repeated observations of the same targets and traditional focus on the sharp-lined
stars, for which the spectrum variations are much harder to detect. Although
the reality of chemical spots in HgMn spots has been convincingly established,
the physical interpretation of this phenomena remains uncertain. A possible
key to understanding the spottedness of HgMn stars is provided by our follow up
study of \aand\ (Kochukhov et al. \cite{KAG07}). Using exceedingly high-quality
spectra ($S/N>1000$), we have reconstructed a series of Hg Doppler images,
covering the time span of 7 years.  The resulting chemical maps for the three
different epochs are shown in Fig.~\ref{fig3}. It is evident that the Hg spot
topology changes slowly over time. This is the first ever observation of
evolving chemical structure at the surface of any star. Our breakthrough
finding suggests that the process responsible for the Hg spot formation in
\aand\ is fundamentally different from the interplay of strong magnetic field
and radiative
diffusion acting in Ap/Bp stars. In contrast to the latter objects, in HgMn
stars we find an unstable surface structure, created without magnetic field
participation. 

Kochukhov et al. (\cite{KAG07}) have suggested that theoretically expected
(Michaud et al. \cite{MRC74}) fine balance of the gravitational force and radiative
pressure in the Hg-rich layer in the upper atmosphere of \aand\ can be
disrupted by hydrodynamical instabilities or by the time-dependent diffusion
effects (Alecian \cite{A98}). Thus, the structure of mercury clouds is governed by a stochastic
process, similar to the cloud weather in planetary atmospheres. Detailed theoretical
calculations are however required to assess feasibility of this hypothesis.

The discovery of mercury cloud weather in \aand\ and detection of chemical inhomogeneities in
other HgMn stars has important implications for our understanding of the
properties of various classes of spotted CP stars. On the one hand, chemical
weather effects can be present in magnetic stars, partly explaining the
observed diversity of abundance distributions. On the other hand, large
discrepancies of the heavy element abundances in HgMn stars of the same mass
and age (e.g., Smith \cite{S97}) and in the nearly equal components of the
close binary systems (e.g., Zverko et al. \cite{ZZK97}; Catanzaro \& Leone
\cite{CL06}) can be attributed to
the slow variation of the surface abundances under the influence of the same
process that acts in \aand. If different stars are observed at different phases
of their long-term ``weather cycle'', the inferred star-to-star scatter in the
average abundances may be essentially random or only weakly related to the
fundamental stellar properties. In this context the discovery of spotted
distribution for precisely those elements (Hg, Pt, Y, Zr) which show the most
extreme abundance anomalies {\it and} large star-to-star scatter is not 
coincidental.

\end{document}